\documentclass[aps,prd,twocolumn,amsmath]{revtex4-1}  %
\usepackage[utf8]{inputenc} 
\usepackage{color,epsfig,graphicx,amssymb,float,multirow,xspace,times} 
\definecolor{dblue}{rgb}{0.00,0.00,0.75}
\usepackage[colorlinks,urlcolor=dblue,linkcolor=dblue,citecolor=dblue]{hyperref} 

\begin{document}   

      \title{Contributions of $\rho(770,1450)\to \omega\pi$ for the Cabibbo-favored $D \to h\omega\pi$ decays} 
                               
\author{Wen-Fei Wang$^{1}$}\email{wfwang@sxu.edu.cn}                
\author{Jiao-Yuan Xu$^{1}$}    
\author{Si-Hong Zhou$^{2}$}\email{shzhou@imu.edu.cn}
\author{Pan-Pan Shi$^{3}$}\email{Panpan.Shi@ific.uv.es}

\affiliation{$^1$Institute of Theoretical Physics and \\
                           State Key Laboratory of Quantum Optics and Quantum Optics Devices, \\
                           Shanxi University, Taiyuan, Shanxi 030006, China \\
                    $^2$Mongolia Key Laboratory of Microscale Physics and Atom Innovation,  \\
                            School of Physical Science and Technology, \\
                           Inner Mongolia University, Hohhot 010021, China\\                
                    $^3$Departamento de F\'{\i}sica Te\'orica and IFIC, Centro Mixto Universidad de \\
	                  Valencia-CSIC Institutos de Investigaci\'on de Paterna, Aptdo.~22085, 46071 Valencia, Spain      
	       }  

\date{\today}

\begin{abstract}
Recently, the BESIII Collaboration has observed the three-body decays $D_s^+\to \eta \omega\pi^+$, 
$D^+\to K^0_S\pi^+\omega$ and $D^0\to K^-\pi^+\omega$. In this work, we investigate the contributions 
of the subprocesses $\rho^+\to \omega\pi^+$ in these Cabibbo-favored decays $D \to h\omega\pi$, with 
$\rho^+= \{\rho(770)^+, \rho(1450)^+, \rho(770)^+\&\rho(1450)^+\}$ and $h=\{ \eta, K^0_S, K^-\}$, by introducing 
these subprocesses into the decay amplitudes of relevant decay processes via the vector form factor $F_{\omega\pi}$ 
which has measured in the related $\tau$ and $e^+e^-$ processes; we provide the first theoretical predictions for the 
branching fractions of the quasi-two-body decays $D_s^+\to\eta[\rho^+\to]\omega\pi^+$, 
$D^+\to K^0_S[\rho^+\to]\omega\pi^+$ and $D^0\to K^-[\rho^+\to]\omega\pi^+$. 
Our findings reveal that the contributions from the subprocess $\rho(770)^+\to\omega\pi^+$ are significant in these observed 
three-body decays $D_s^+\to\eta \omega\pi^+$, $D^+\to K^0_S \omega\pi^+$ and $D^0\to K^- \omega\pi^+$, 
notwithstanding the contributions originating from the Breit-Wigner tail effect of $\rho(770)^+$. 
The numerical results of this study suggest that the dominant resonance contributions for the three-body decays 
$D_s^+\to\eta \omega\pi^+$ and $D^+\to K^0_S \omega\pi^+$ are originated from the $P$-wave intermediate states 
$\rho(770)^+$, $\rho(1450)^+$ and their interference effects. 
\end{abstract}

\maketitle

\section{Introduction}
\label{sec-intro}  

The three-body decay $D_s^+\to \eta \omega\pi^+$ has been observed for the first time by BESIII 
Collaboration recently, with its total branching fraction $\mathcal{B}=(0.54\pm0.12_{\rm stat}\pm0.04_{\rm syst})\%$~\cite{prd107.052010}. This numerical result aligns an earlier measurement by the CLEO Collaboration, which reported a 
branching fraction of $(0.85\pm0.54_{\rm stat}\pm0.06_{\rm syst})\%$ in Ref.~\cite{prd80.051102} for the same 
decay channel, but the BESIII measurement offers significantly improved precision.  Given the quark structure of the 
initial and the final states  in this $D_s$ decay process, the Cabibbo-favored transition $c\to s$ along with the 
$W^+\to u\bar d$ will be the dominated process at quark level.  The prospective intermediate states for this process 
could be the resonances $a_0(980)^+$, $\rho(770)^+$, $b_1(1235)^+$, and $\omega(1420)$, etc., and their excited 
states which will decay into the $\pi^+ \eta$, $\omega\pi^+$ and $\omega\eta$ pairs in the final state, 
respectively~\cite{PDG-2024}.  The combination of $\pi^+$ and $\omega\eta$ with the intermediate resonance 
$\omega(1420)$ means a pure annihilation process for this $D_s^+$ decay. The union of $\omega$ 
and $a_0(980)^+\to \pi^+ \eta$ for this decay is very similar to the decay process 
$D_s^+\to \pi^0[a_0(980)^+\to]\pi^+ \eta$, which has been measured by BESIII~\cite{prl123.112001} and 
the branching fraction was found to be consistent with the triangle rescattering processes via the intermediate 
processes $D_s^+\to \eta^{(\prime)}\rho^+$, $D_s^+\to K^* \bar{K}$ and $D_s^+\to K \bar{K^*}$ \cite{epjc80-895,plb803-135279,epjc80-1041,prd103.116016,prd109.076027,epjc81-1093}. But the branching 
fraction for $D_s^+\to \omega a_0(980)^+$ was found to be less than $3.4\times 10^{-4}$ in Ref.~\cite{epjc81-1093}, 
due to the cancellation of the rescattering effects and the suppressed short-distance $W$ annihilation contribution.
Consequently, the dominated contributions for the decay $D_s^+\to  \eta\omega\pi^+$ are expected to come from 
the combination of $\eta$ and the resonances which will decay into $\omega\pi^+$ pair in the final state.

The light meson pair of $\omega\pi$ which originating from the weak current of the matrix element 
$\langle\omega\pi|V^\mu\!-\!A^\mu|0\rangle$ in the three-body hadronic $D$ and $B$ meson decays is related 
to the processes $\tau\to \omega\pi\nu_\tau$ and $e^+e^-\to \omega\pi$.  The $G$-parity conservation requires 
that the  $\omega\pi$ pair is determined mainly by the vector part $V^\mu$ of the weak current~\cite{zpc57-495}, 
which is associated with the $\rho$ family resonances as the intermediate states via the $P$-wave transition 
amplitudes~\cite{prd86.037302}. The axial-vector part $A^\mu$ of the weak current for $\omega\pi$ system 
goes to the resonance $b_1(1235)$ and its excited states.  The state $b_1(1235)$, with the quantum numbers 
$J^{PG}=1^{++}$, is related to the second-class weak current classified in terms of parity and $G$ 
parity~\cite{pr112.1375}.  The second-class weak current has been searched in $\tau$ decays in recent years, 
but no evidence has been observed~\cite{ar0807.4900,npb218-110}. The decay amplitude for the 
$b_1(1235)\to \omega\pi$ subprocess through $S$- and $D$-waves in relevant processes is proportional to the 
mass difference between $u$ and $d$ quarks~\cite{PDG-2024}, which makes the contribution of $\omega\pi$ 
from the axial-vector resonance $b_1(1235)$ accompanied by the second-class current highly suppressed.

The ordinary decay mode for $\rho(770)\to \omega\pi$ is not allowed because of the pole mass of $\rho(770)$, 
which is apparently below the threshold of the $\omega\pi$ pair~\cite{PDG-2024}. But the Breit-Wigner (BW) 
\cite{BW-model} tail effect of $\rho(770)$, also known as the virtual contribution~\cite{plb25-294,Dalitz62,
prd94.072001,plb791-342}, was found indispensable to the production of $\omega\pi$ pair in the processes of 
$\tau \to \omega \pi \nu_\tau$~\cite{plb185-223,prd61.072003,rmp78.1043,prl103.041802}  
and $e^+e^- \to\omega\pi^0$~\cite{plb174-453,plb486-29,plb562-173,plb669-223,JETPL94-734,prd88.054013,
prd94.112001,prd96.092009,plb813-136059,2309.00280}. The excited state $\rho(1450)$ has been observed to 
decay into $\omega\pi$ pair in $B$ meson decays by the CLEO, Babar and Belle collaborations~\cite{prd64.092001,
prd74.012001,prd92.012013}. This state has been suggested as a $2S$-hybrid mixture in Ref.~\cite{prd60.114011} 
in view of its decay characters~\cite{npb443-233,prd52.1706,prd56.1584}, but its mass is consistent with 
the $2S$ excitation of  $\rho(770)$~\cite{prd55.4157}. The further investigation of the interference between the 
$\rho(1450)$ and its ground state will provide deeper insights into its nature.

\begin{figure}[tbp]  
\centerline{\epsfxsize=5.5cm \epsffile{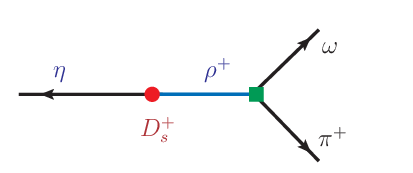}}
\caption{Schematic view of the cascade decay $D_s^+\to \eta  \rho^+\to \eta \omega\pi^+$, 
              where $\rho^+$ stands for the intermediate states $\rho(770,1450)^+$ which decay into $\omega\pi^+$  
              in this work. }
\label{fig-1}
\end{figure}

In addition to data for the $D_s^+\to \eta \omega\pi^+$ decay, the absolute branching fractions of the decays 
$D^0\to K^-\pi^+\omega$, $D^0\to K^0_S\pi^0\omega$, and $D^+\to K^0_S\pi^+\omega$ have been 
determined by BESIII Collaboration recently in Ref.~\cite{prd105.032009}, with the results  to be 
$(3.392 \pm 0.044_{\rm stat} \pm 0.085_{\rm syst})\%$, $(0.848 \pm 0.046_{\rm stat} \pm 0.031_{\rm syst})\%$,
and $(0.707 \pm 0.041_{\rm stat} \pm 0.029_{\rm syst})\%$, respectively.  To better understand these experimental 
results, we shall study the Cabibbo-favored three-body decays $D_s^+\to  \eta\omega\pi^+$, 
$D^+\to  K^0_S\omega\pi^+$ and $D^0\to  K^-\omega\pi^+$ in this work, where the $\omega\pi$ pair is attributed 
to the decays of intermediate states $\rho(770)^+$ and $\rho(1450)^+$.  The schematic 
diagram for the quasi-two body decay $D_s^+\to \eta \rho^+\to \eta\omega\pi^+$ is shown in Fig.~\ref{fig-1}. 
In its rest frame, the state $D_s^+$ will decay into the intermediate resonance $\rho^+$ along with the the bachelor 
state $\eta$, and then the $\rho^+$ decays into $\omega$ and $\pi^+$ via the strong interaction. The similar pattern 
will arise in $D^{+}$ and $D^0$ decay processes, with the $\eta$ replaced by $K^0_S$ and $K^-$, respectively. 
The subprocesses $\rho(770,1450)^+\to \omega\pi^+$ in these decays will be introduced into their decay amplitudes 
in the isobar formalism~\cite{pr135.B551,pr166.1731,prd11.3165} via the vector form factor $F_{\omega\pi}$. 
This form factor has been measured in the related processes of $\tau$ decay and $e^+e^-$ annihilation. 

The contributions of $\rho(770,1450)\to \omega\pi$ for the three-body decays $B\to\bar{D}^{(*)} \omega\pi$ were studied 
in Ref.~\cite{JHEP2401-047} very recently. The $\rho$ family resonance contributions of kaon pair have been explored 
in Refs.~\cite{prd101.111901,prd103.056021,prd103.016002,cpc46-053104,prd107.116023,prd110.056001,prd109.116009} 
and in Refs.~\cite{prd85.092016,prd93.052018,prd103.114028,prd104.116019} for the three-body $B$ and $D$ meson 
decays, respectively. In Ref.~\cite{2309.00280}, four resonances $\rho(770)$, $\rho(1450)$, $\rho(1700)$ and $\rho(2150)$ 
have been employed to parametrize the related transition form factor for $\rho\to\omega\pi$ for the process 
$e^+e^- \to \omega\pi^0 \to \pi^+\pi^-\pi^0\pi^0$ in the energy range $1.05$-$2.00$ GeV by SND Collaboration. 
But we will leave the contributions of $\rho(1700, 2150)\to\omega\pi$ to future studies for the relevant  decays, 
in view of that the masses of $\rho(1700, 2150)$ are very close to or even beyond the masses of the initial $D^{+,0}$ 
and $D^+_{s}$ mesons, the contributions of $\omega\pi$ from $\rho(1700, 2150)$ are unimportant 
comparing with that from $\rho(770)$ and $\rho(1450)$~\cite{2309.00280}, and in addition, the excited $\rho$ states 
around $2$ GeV are not well understood~\cite{plb813-136059,prd105.074035}.

Due to the $c$-quark mass,  the heavy quark expansion tools and the factorization approaches, which have been 
successfully used in hadronic $B$ meson decays for decades, encounter significant challenges when applied to the
two-body or three-body hadronic $D$ meson decays. In this context, model independent methods, 
such as the topological-diagram 
approach~\cite{prd85.054014,prd85.034036,prd86.014014,prd93.114010,prd100.093002,prd104.073003,prd110.094052} 
and factorization-assisted topological-amplitude approach~\cite{prd86.036012,prd89.054006,adv-7627308,ar2503.18593} have been adopted in various $D$ decay studies. In Ref.~\cite{prd104.116019},  we construct a theoretical framework for 
quasi-two-body $D$ meson decays with the help of electromagnetic form factors, with which we studied the contributions 
of $\rho(770,1450)\to K\bar{K}$ for the three-body $D$ decays within the flavour \emph{SU}(3) symmetry. In this work 
we adopt the method in~\cite{prd104.116019} and investigate the concerned decays within the quasi-two-body 
framework~\cite{1605.03889,prd96.113003,prd104.116019}, while neglect the interaction between $\omega\pi$ system 
and the corresponding bachelor state in relevant decay processes.

This paper is organized as follows. In Sec.~\ref{sec:2}, we take $D^0 \to K^-[\rho(770)^+\to] \omega\pi^+$ as an 
example to derive the differential branching fractions for the quasi-two-body $D$ meson decay processes.  
In Sec.~\ref{sec-res}, we present our numerical results for the branching fractions of the quasi-two-body decays 
$D_s^+\to\eta[\rho^+\to]\omega\pi^+$, $D^+\to K^0_S[\rho^+\to]\omega\pi^+$ and $D^0\to K^-[\rho^+\to]\omega\pi^+$,
along with some necessary discussions. Summary of this work is given in Sec.~\ref{sec-sum}.

\section{Differential branching fractions}\label{sec:2}
\label{sec-frame}  
\begin{figure}[tbp]  
\centerline{\epsfxsize=6.5cm \epsffile{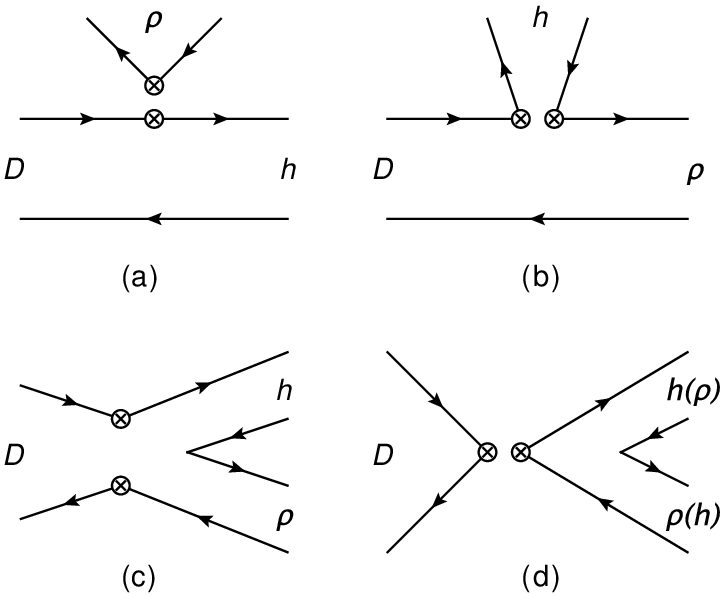}}
\caption{Typical topological diagrams for the concerned decays at quark level, 
              where $\rho$ stands for the intermediate states $\rho(770,1450)^+$, 
              $h$ for $\eta, K^0_S$ or $K^-$, 
              and the symbol $\otimes$ stands for the weak vertex
              in this work. }
\label{fig-ql}
\end{figure}

In this section, we take the decay $D^0 \to K^-[\rho(770)^+\to] \omega\pi^+$ as an example to derive the differential 
branching fraction for the quasi-two-body $D$ meson decay processes. If the subprocess $\rho(770)^+\to \omega\pi^+$ 
was shrunk to the meson $\rho(770)^+$, we will have the two-body decay $D^0 \to K^-\rho(770)^+$. The related 
effective weak Hamiltonian for $D$ decays is written as~\cite{rmp68.1125}
\begin{eqnarray}
\mathcal{H}_{\rm eff} = \frac{G_F}{\sqrt2}&\bigg[& \sum_{q=d,s} \lambda_q (C_1 O_1+ C_2 O_2)                        \nonumber \\
                                                                   & - & \lambda_b \sum^{6}_{i=3}C_i O_i  -  \lambda_b C_{8g} O_{8g}\bigg],
\label{eq-hamilton}
\end{eqnarray}
where the Fermi coupling constant $G_F=1.1663788(6)\times10^{-5}$ GeV$^{-2}$~\cite{PDG-2024}, the product 
of the Cabbibo-Kobayashi-Maskawa (CKM)~\cite{Cabibbo,Kobayashi} matrix elements $\lambda_q=V^*_{cq}V_{uq}$ 
and $\lambda_b=V^*_{cb}V_{ub}$, the $C$'s are Wilson coefficients at scale $\mu$, and the $O_1$ and $O_2$ are 
current-current operators, $O_3$-$O_4$ are QCD penguin operators, and $O_{8g}$ is chromomagnetic dipole operator.
Then the total decay amplitude for the process of a $D$ meson decays into a pseudoscalar ($P$) plus a vector ($V$) can 
be described with typical topological diagram amplitudes $T_{P,V}, C_{P,V}, E_{P,V}$ and $A_{P,V}$ according 
to the diagrams shown in Fig.~\ref{fig-ql}, as well as 
additional penguin amplitudes in the factorization-assisted topological-amplitude approach~\cite{prd89.054006} and 
the topological-diagram approach~\cite{prd85.034036,prd93.114010,prd100.093002,prd104.073003}. One is referred to 
the Refs.~\cite{prd89.054006,prd85.034036,prd93.114010,prd100.093002,prd104.073003} for the detailed discussions 
on these the topological diagram amplitudes.

The decay amplitude for $D^0 \to K^-\rho(770)^+$ is dominated by the color-favored tree amplitude $T_P^{2B}$ with 
the $D^0 \to K^-$ transition as shown in Fig.~\ref{fig-ql}-(a), where the superscript $2B$ stands for a two-body decay process. We have the amplitude $T_P^{2B}$ as~\cite{prd89.054006,prd93.114010,prd100.093002}
\begin{eqnarray} 
  T_P^{2B} &=& \frac{G_F}{\sqrt2} V^*_{cs}V_{ud} \left[ \frac{C_1}{3}+C_2 \right] f_\rho m_\rho 
                    F_1^{D\to \bar K}(m^2_\rho) 2 \epsilon_\rho\!\cdot\! p_D\;   \nonumber\\  
                   &=& \mathcal{M}_T^{2B} \epsilon_\rho\!\cdot\! p_D,
  \label{def-Tp}
\end{eqnarray}  
where the subscript $\rho$ stands for the resonance $\rho(770)$ here, the $F_1^{D\to \bar K}(m^2_\rho)$ is the transition 
form factor for $D^0 \to K^-$ process. Beyond the dominated amplitude $T_P^{2B}$, there is a $W$-exchange 
nonfactorizable contribution from the amplitude $E^{2B}_V$ from the topological diagram Fig.~\ref{fig-ql}-(c), 
which can be parametrized as the form 
$E^{2B}_V=\mathcal{M}^{2B}_E\epsilon_\rho\!\cdot\! p_D$, and we have~\cite{prd89.054006} 
\begin{eqnarray}
   \mathcal{M}^{2B}_E&=& \frac{G_F}{\sqrt2} V^*_{cs}V_{ud} C_2 \chi^E_q e^{i \phi^E_q} f_D m_D  
                          \left[ f_K/f_\pi \right], 
    \label{eq-M2B}
\end{eqnarray}
in the factorization-assisted topological-amplitude approach, with $\chi^E_q=0.25$ and 
$\phi^E_q=1.73$~\cite{prd89.054006} are two parameters which characterize the strength and strong phase of the
corresponding amplitude, and the decay constants $f_\rho$, $f_D$, $f_K$ and $f_\pi$ for the $\rho(770)$, $D^0$, kaon 
and pion, respectively. 
In addition to the amplitudes $T_P^{2B}$ and $E^{2B}_V$, we also need~\cite{prd89.054006} 
\begin{eqnarray}
   C^{2B}_V&=&\mathcal{M}^{2B}_C\epsilon_\rho\!\cdot\! p_D           \nonumber\\
                    &=&\frac{G_F}{\sqrt2} V^*_{cs}V_{ud} \left[C_1+C_2(1/3+\chi^C_V e^{i \phi^C_V})\right]       \nonumber\\
                    &\times&f_K m_\rho A^{D\to \rho}_0(m^2_K)2\,\epsilon_\rho\!\cdot\! p_D, \\
    A^{2B}    &=&\mathcal{M}^{2B}_A\epsilon_\rho\!\cdot\! p_D          \nonumber\\
                     &=&\frac{G_F}{\sqrt2} V^*_{cs}V_{ud} C_1 \chi^A_q e^{i \phi^A_q} f_D m_D  
                          \left[ f_{\eta^{(\prime)}}/f_\pi \right]\epsilon_\rho\!\cdot\! p_D,\;\; 
    \label{eq-A2B}
\end{eqnarray}
in the numerical calculations of the decays $D^+\to K^0_S\rho(770)^+$ and $D_s^+\to\eta \rho(770)^+$ 
according to the topological diagrams Fig.~\ref{fig-ql}-(b) and -(d), respectively. 
With the parameters $\chi^C_V=-0.53$, $\phi^C_V=-0.25$, $\chi^A_q=0.11$ and $\phi^A_q=-0.35$ 
found in~\cite{prd89.054006}.

For the two-body $D^0 \to K^-\rho(770)^+$ decay, the Lorentz-invariant total decay amplitude is 
\begin{equation}
     \mathcal{A}^{2B}=\mathcal{M}^{2B} \epsilon_\rho\!\cdot\! p_D 
                                    =(\mathcal{M}_T^{2B}+\mathcal{M}^{2B}_E)\epsilon_\rho\!\cdot\! p_D.
\end{equation}
 Utilizing the partial decay rate
\begin{equation}
        d\Gamma=\frac{1}{32\pi^2}|\mathcal{A}|^2 \frac{\big|\overrightarrow{q}\big|}{m^2_D} d\Omega
\end{equation}
in  {\it Review of Particle Physics}~\cite{PDG-2024}, along with the formula
\begin{eqnarray}
  \sum_{\lambda=0,\pm} \varepsilon^{*\mu}{(p,\lambda)}\varepsilon^{\nu}{(p,\lambda)}
                                     =-g^{\mu\nu} + \frac{p^\mu p^\nu}{p^2},
  \label{def-sumofpol}
\end{eqnarray}
it is trivial to get the partial decay width
\begin{eqnarray} 
   \Gamma(D^0 \to K^-\rho(770)^+)=\frac{\big| \overrightarrow{q}\big|^3}{8\pi m^2_\rho}\big|\mathcal{M}^{2B}\big|^2\!.
   \label{def-dew-2B}       
\end{eqnarray}
The magnitude of the momentum $|\overrightarrow{q}|$ for $\rho(770)^+$ or $K^-$ is
\begin{eqnarray}
   &&\left| \overrightarrow{q} \right|=   \nonumber\\
   &&\;\frac{1}{2 m_D} \sqrt{\left[m^2_{D}-(m_\rho+m_{K})^2\right]\left[m^2_{D}-(m_\rho-m_{K})^2\right]},\;\;\;
    \label{def-2B-p}
\end{eqnarray}  
in the rest frame of $D^0$ meson, and the $m_{i}$'s (with $i=\{D^0, \rho(770)^+, K^-\}$) are the masses for relevant 
particles above.

By connecting subprocess $\rho(770)^+\to \pi^+\pi^0$ and the two-body mode $D^0 \to K^-\rho(770)^+$ together, 
we will get the quasi-two-body decay $D^0 \to K^-\rho(770)^+\to K^-  \pi^+\pi^0$. Its decay amplitude can be written 
as~\cite{prd104.116019} 
\begin{equation}
     \mathcal{A}^{Q2B}=\mathcal{M}^{Q2B} \epsilon_\rho\!\cdot\! p_D
                            \frac{1}{\mathcal{D}^\rho_{\rm BW}} g_{\rho\pi\pi}  \epsilon_\rho\cdot(p_{\pi^+}-p_{\pi^0}).
                               \label{def-A-Q2B}
\end{equation}
Where the BW denominator $\mathcal{D}^\rho_{\rm BW}=m^2_\rho-s-im_\rho\Gamma_\rho(s)$ for the 
$\rho(770)^+$ propagator, and the related $s$-dependent width is     
\begin{eqnarray}\label{def-width}
 \Gamma_{\rho}(s)
             =\Gamma_\rho\frac{m_\rho}{\sqrt s} \frac{ \left| \overrightarrow{q_\pi} \right|^3}{ \left| \overrightarrow{q_{\pi0}}\right|^3} 
                X^2(\left| \overrightarrow{q_\pi} \right| r^\rho_{\rm BW}).
  \label{eq-sdep-Gamma}
\end{eqnarray}
The Blatt-Weisskopf barrier factor for $\rho$ family resonances is given by~\cite{BW-X}  
\begin{eqnarray}
     X(z)=\sqrt{\frac{1+z^2_0}{1+z^2}}\,,
\end{eqnarray}
where $\left| \overrightarrow{q_\pi} \right|=\frac{1}{2}\sqrt{s-4m_\pi^2}$, 
$\left| \overrightarrow{q_{\pi0}}\right|$ is $\left| \overrightarrow{q_\pi} \right|$ at $s=m^2_\rho$, and the barrier radius is $r^\rho_{\rm BW}=1.5$ GeV$^{-1}$~\cite{prd63.092001}.
One should note that the amplitude $\mathcal{M}^{Q2B}$ could be obtained from the related amplitude 
$\mathcal{M}^{2B}$ with the replacement $m_\rho\to \sqrt{s}$. With the help of Eq.~(\ref{def-sumofpol}), the decay 
amplitude $\mathcal{A}^{Q2B}$ of Eq.~(\ref{def-A-Q2B}) is reduced to
\begin{equation}
     \mathcal{A}^{Q2B}=\mathcal{M}^{Q2B}  \frac{g_{\rho\pi\pi}}{\mathcal{D}^\rho_{\rm BW}} 
                                        \big|\overrightarrow{p_1}\big|\cdot\big|\overrightarrow{p_3}\big|\cos\theta, 
                 \label{def-AofQ2B}
\end{equation}
where 
\begin{eqnarray} 
             \big|\overrightarrow{p_1}\big| &=&  \frac{1}{2}\sqrt{s-4m_\pi^2}, \nonumber\\
             \big|\overrightarrow{p_3}\big| &=&  \frac{1}{2 \sqrt{s}} 
                                 \sqrt{\left[m^2_{D}-(\sqrt{s}+m_{K})^2\right]\left[m^2_{D}-(\sqrt{s}-m_{K})^2\right]},\nonumber\\
\end{eqnarray}
are the momenta for final states $\pi^+$ and $K^-$, respectively, in the rest frame of the pion pair in the 
$D^0 \to K^-  \pi^+\pi^0$ decay, and $\theta$ is the angle between $\pi^0$ and $K^-$ in the same frame for pion pair. 
After the integration of $\cos\theta$, it's trivial to arrive the partial decay width~\cite{PDG-2024}
\begin{eqnarray} 
   \frac{d\Gamma}{ds}=\frac{\big|\overrightarrow{p_1}\big|^3 \big|\overrightarrow{p_3}\big|^3}{48\pi^3 m_D^3}
                \left|\mathcal{M}^{Q2B}  \frac{g_{\rho\pi\pi}}{\mathcal{D}^\rho_{\rm BW}} \right|^2\!,
   \label{def-dew-Q2B}       
\end{eqnarray}
for the quasi-two-body decay $D^0 \to K^-[\rho(770)^+\to] \pi^+\pi^0$. We can further define the pion 
electromagnetic form factor of the $\rho(770)$ component as~\cite{epjc39-41,prd81-094014}   
\begin{equation}
    F_\pi(s)=\frac{f_\rho g_{\rho\pi\pi}}{\sqrt2\, m_\rho}\frac{m^2_\rho}{\mathcal{D}^\rho_{\rm BW}}, 
\end{equation}
and rewrite the Eq.~(\ref{def-dew-Q2B}) as 
\begin{eqnarray} 
   \frac{d\Gamma}{ds}=\frac{\big|\overrightarrow{p_1}\big|^3 \big|\overrightarrow{p_3}\big|^3}{24\pi^3 m_D^3}
                \left|\mathcal{M}^{Q2B}\right|^2_{f_\rho m_\rho\to F_\pi}\!,
   \label{PDW-Q2B-pi}       
\end{eqnarray}
which is the same expression as it in Ref.~\cite{prd104.116019}.

To calculate the quasi-two-body decay $D^0 \to K^-[\rho(770)^+\to] \omega\pi^+$, we introduce the effective Lagrangian~~\cite{plb66-165,prd30.594,prd46.1195}
\begin{eqnarray}
       \mathcal{L}_{\rho\omega\pi}=g_{\rho\omega\pi}
                     \epsilon_{\mu\nu\alpha\beta}\partial^{\mu}\rho^{\nu}\partial^{\alpha}\omega^{\beta}\pi
        \label{eq-Lagrangian}
\end{eqnarray}
to describe the $\rho$ and $\omega\pi$ coupling. 
The related form factor $F_{\omega\pi}(s)$ is expressed as~\cite{epja38-331,prd92.014014,2307.10357}
\begin{eqnarray}   
      \langle \omega(p_a,\lambda)\pi(p_b)| j_\mu(0) | 0 \rangle      
            = i  \epsilon_{\mu\nu\alpha\beta}\varepsilon^{\nu}(p_a, \lambda)p_b^\alpha p^\beta F_{\omega\pi}(s),\;\;
  \label{eq-def-FOpi}  
\end{eqnarray}
where $j_\mu$ is the isovector part of the electromagnetic current, $\lambda$ and $\varepsilon$ are the polarization 
and polarization vector for $\omega$, respectively, $p_a(p_b)$ is the momentum for $\omega(\pi)$, and the momentum
$p=p_a+p_b$ for the resonance $\rho(770)^+$. The form factor $F_{\omega\pi}(s)$ in the vector meson dominance 
model is parametrized as~\cite{plb486-29,prd88.054013,prd94.112001,ppnp120-103884}
\begin{eqnarray}
	F_{\omega\pi}(s) = \frac{g_{\rho\omega\pi}}{f_\rho} \sum\limits_{\rho_i}  
	               \frac{A_i e^{i \phi_i} m_{\rho_i}^2 }{D_{\rho_i}(s)}\,,
  \label{exp-formfactor}
\end{eqnarray}   
where the summation is over the isovector resonances $\rho_i=\{\rho(770), \rho(1450), \rho(1700), ...\}$ in $\rho$ 
family, with $m_{\rho_i}$ their masses. The $A_i$ and $\phi_i$ are the weights and phases for these resonances, 
respectively, and we can assign $A=1$ and $\phi=0$ for $\rho(770)$. Technically, the contributions from the 
excitations of $\omega(782)$ should also be include in Eq.~(\ref{exp-formfactor}), but their weights were found to 
be negligibly small~\cite{jpg36-085008}. 

The decay amplitude for the quasi-two-body process $D^0 \to K^-[\rho(770)^+\to] \omega\pi^+$ is primarily written 
as
\begin{equation}
     \mathcal{A}_{\omega\pi}=\mathcal{M}_{\omega\pi} \epsilon_\rho\!\cdot\! p_D
                            \frac{1}{\mathcal{D}_{\rm BW}} g_{\rho\omega\pi} 
                            \epsilon_{\mu\nu\alpha\beta} \epsilon^\mu_\rho\varepsilon^{\nu}_\omega p_b^\alpha p^\beta.
                               \label{def-Op-Q2B}
\end{equation}
With the help of the relation~\cite{JHEP2401-047}
\begin{eqnarray}
 &&\sum_{\lambda=0,\pm}\vert\epsilon_{\mu\nu\alpha\beta}p_{3}^\mu\varepsilon^{\nu}(p_\omega, \lambda)
          p_\pi^\alpha p^\beta\vert^2  \nonumber\\
  && \qquad   = s\,\vert\overrightarrow{p_\omega}\vert^2 \vert\overrightarrow{p_3}\vert^2 (1-\cos^2{\theta}), 
\end{eqnarray}
we have the partial decay width~\cite{PDG-2024}
\begin{eqnarray} 
   \frac{d\Gamma}{ds}=\frac{s\big|\overrightarrow{p_\omega}\big|^3 \big|\overrightarrow{p_3}\big|^3}{96\pi^3 m_D^3}
                \left|\mathcal{M}_{\omega\pi} \frac{g_{\rho\omega\pi}}{\mathcal{D}_{\rm BW}} \right|^2\!,
   \label{PDW-Q2B-Op}       
\end{eqnarray}
for the quasi-two-body decay $D^0 \to K^-[\rho(770)^+\to] \omega\pi^+$ after the integration of $\cos\theta$. 
Then we can define an auxiliary electromagnetic form factor
\begin{eqnarray} 
      f_{\omega\pi}=f^2_\rho/m_\rho F_{\omega\pi}=f_\rho m_\rho g_{\rho\omega\pi}/{\mathcal{D}_{\rm BW}},
\end{eqnarray}
and rewrite the Eq.~(\ref{PDW-Q2B-Op}) as 
\begin{eqnarray} 
   \frac{d\Gamma}{ds}=\frac{s\big|\overrightarrow{p_\omega}\big|^3 \big|\overrightarrow{p_3}\big|^3}{96\pi^3 m_D^3}
                \left|\mathcal{M}_{\omega\pi}\right|^2_{f_\rho m_\rho\to f_{\omega\pi}}\!.
   \label{PDW-Q2B-piO}       
\end{eqnarray}
One should note that this expression is a little different from the differential branching fraction in 
Ref.~\cite{JHEP2401-047} for $B\to\bar{D}^{(*)} \omega\pi$ decays.  This discrepancy arises from the different 
definitions for the quasi-two-body decay amplitudes between the perturbative QCD approach~\cite{JHEP2401-047} 
and the present paper.

We need to stress that the $\mathcal{D}_{\rm BW}=m^2_\rho-s-i\sqrt{s}\,\Gamma_\rho(s)$ for the $\rho(770)^+$ in 
Eq.~(\ref{def-Op-Q2B}) is different from the ${\mathcal{D}^\rho_{\rm BW}}$ in Eq.~(\ref{def-A-Q2B}). The former 
has the expression~\cite{2309.00280,prd88.054013}
\begin{eqnarray}
 \Gamma_{\rho(770)}(s) &=& \Gamma_{\rho(770)}\frac{m^2_{\rho(770)}}{s}
    \left(\frac{\vert\overrightarrow{p_1}\vert} {\vert\overrightarrow{p_1}\vert_{s=m^2_{\rho(770)} } }\right)^{3}    \nonumber\\
  &+&\frac{g_{\rho\omega\pi}^2}{12\pi} \vert\overrightarrow{p_\omega}\vert^3,
\label{eq-Gm-R770}
\end{eqnarray}
for the energy-dependent width for the resonance $\rho(770)$, with the 
\begin{eqnarray}    
     \vert\overrightarrow{p_\omega}\vert
                       =\frac{1}{2\sqrt s} \sqrt{\left[s-(m_\omega+m_{\pi})^2\right]\left[s-(m_\omega-m_{\pi})^2\right]}\,.
     \label{def-qO}           
\end{eqnarray}  
In addition to the $\Gamma_{\rho(770)}(s)$, we employ the expression
\begin{eqnarray}    
    \Gamma_{\rho(1450)}(s) =\Gamma_{\rho(1450)} \Bigl[ {\mathcal B}_{\rho(1450) \to \omega\pi}  
          \Bigl(\frac {q_{\omega}(s)} {q_{\omega}(m^2_{\rho(1450)})}\Bigr)^3\;         \nonumber\\
     \quad\qquad   +(1-{\mathcal B}_{\rho(1450) \to \omega\pi})
           \frac {m^2_{\rho(1450)}} {s}  \Bigl(\frac {q_{\pi}(s)} {q_{\pi}(m^2_{\rho(1450)})}\Bigr)^3\Bigr]\;\;
 \label{eq-Gm-R1450}
\end{eqnarray}  
for the excited resonance $\rho(1450)$ for its energy-dependent width as it was adopted in Ref.~\cite{plb562-173} 
for the process $e^+ e^-\to\omega\pi^0\to\pi^0\pi^0\gamma$ by CMD-2 Collaboration, where 
${\mathcal B}_{\rho(1450) \to \omega\pi}$ is the branching ratio of the $\rho(1450) \to \omega\pi$ decay, 
$\Gamma_{\rho(770)}$ and $\Gamma_{\rho(1450)}$ are the full widths for $\rho(770)$ and $\rho(1450)$, 
respectively. 

\section{Numerical results and Discussions} 
\label{sec-res}
The key input in this work is the coupling constant $g_{\rho\omega\pi}$
in Eq.~(\ref{exp-formfactor}) for the subprocess $\rho(770)\to\omega\pi$; its value can be estimated with the 
relation $g_{\rho\omega\pi}\approx3g_{\rho\pi\pi}^2/(8\pi^2F_\pi)$~\cite{prc83.048201} with $F_\pi=92$ 
MeV~\cite{PDG-2024} and can also be calculated from the decay width of $\omega\to\pi^0\gamma$~\cite{PDG-2024}.
In the numerical calculation of this work, we adopt the value $g_{\rho\omega\pi}=16.0\pm2.0$ 
GeV$^{-1}$~\cite{JHEP2401-047} by taking into account the corresponding fitted and theoretical values in 
Refs.~\cite{prd94.112001,prd86.057301,prd61.072003,JETPL94-734,prd55.249,prd77.113011,prd86.037302,prd88.054013} 
for it. In view of the expression of $F_{\omega\pi}(s)$, we have a constraint~\cite{JHEP2401-047} 
\begin{eqnarray}
   A_1= \frac{g_{\rho(1450)\omega\pi}f_{\rho(1450)}m_{\rho(770)}}
                               {g_{\rho(770)\omega\pi}f_{\rho(770)}m_{\rho(1450)}}
   \label{eqn-A1}
\end{eqnarray}
for the weight $A_1$ for subprocess $\rho(1450)\to \omega\pi$ in Eq.~(\ref{exp-formfactor}). With 
the measured result $f^2_{\rho(1450)}\mathcal{B}(\rho(1450)\to\omega\pi)=0.011\pm0.003$ GeV$^2$
\cite{prd64.092001}, we have $A_1=0.171\pm0.036$~\cite{JHEP2401-047}.

\begin{table*}[!]
\centering   
\caption{Masses for relevant states, decay constants for pion and kaon, full widths of $\rho(770)$ and $\rho(1450)$ 
               (in units of GeV) and the CKM matrix elements $|V_{ud}|$ and $|V_{cs}|$ in~\cite{PDG-2024}, 
               along with the decay constants $f_{D}$ and $f_{D_s}$ from~\cite{prd98.074512,epjc80-113}.}
\label{tab1}
\setlength{\tabcolsep}{25pt}
\begin{tabular}{l c c c }\hline\hline
  $m_{D^{\pm}}=1.870$       & $m_{D^{0}}=1.865$           & $m_{D^{\pm}_s}=1.968$       & $m_{\pi^{\pm}}=0.140$           \\
  $m_{\omega}=0.783$         & $m_{K^{\pm}}=0.494$      &  $m_{K^{0}}=0.498$               &  $m_{\eta}=0.548$                     \\  
    $f_{\pi}=0.130$                &      $f_{K}=0.156$                & $f_{D}=0.212$                        &   $f_{D_s}=0.250$\;                       \\  
 \hline                
  \multicolumn{2}{l}{$m_{\rho(770)}=0.775$ }                        &  \multicolumn{2}{r}{$\Gamma_{\rho(770)}=0.147$}        \\
  \multicolumn{2}{l}{$m_{\rho(1450)}=1.465\pm0.025$ }     &  \multicolumn{2}{r}{$\Gamma_{\rho(1450)}=0.400\pm0.060$}    \\
  \multicolumn{2}{l}{$|V_{ud}|=0.97367\pm0.00032$}           &  \multicolumn{2}{r}{$|V_{cs}|=0.975\pm0.006$}                               \\
 \hline\hline   
\end{tabular}
\end{table*}

In the quasi-two body decay $D_s^+\to \eta [\rho^+\to]\omega \pi$, the mixing between $\eta$ and $\eta^{\prime}$ are 
taking into account.  The physical $\eta$ and $\eta^\prime$ states are related to the quark flavor 
basis~\cite{plb449-339,prd58.114006}
\begin{equation}
\left(\begin{array}{c} \eta \\ \eta^{\prime} \end{array} \right)
                      = \left(\begin{array}{cc}  \cos{\phi} & -\sin{\phi} \\ \sin{\phi} & \cos{\phi} \\ \end{array} \right)
 \left(\begin{array}{c} \eta_q \\ \eta_s \end{array} \right),
 \label{mixing-eta}
\end{equation}
the meson $\eta$ is made from $\eta_q=(u\bar u+d\bar d)/\sqrt2$ and $\eta_s=s\bar s$ at quark level in early studies 
with the mixing angle $\phi=39.3^{\circ}\pm1.0^{\circ}$ and the decay constants $f_{\eta_q}=(1.07\pm0.02)f_\pi$ 
and $f_{\eta_s}=(1.34\pm0.06)f_\pi$~\cite{plb449-339,prd58.114006}. 
Recently, the mixing angle $\phi$ has been measured by 
KLOE~\cite{JHEP0907-105,plb648-267}, LHCb~\cite{JHEP1501-024} and BESIII~\cite{prd108.092003,prl122.121801} 
collaborations. In this work, we adopt the angle $\phi=(40.0\pm2.0_{\rm stat}\pm0.6_{\rm syst})^{\circ}$ presented by 
BESIII in Ref.~\cite{prd108.092003} for the $\eta$-$\eta^\prime$ mixing very recently. 

The three-parameter fit formulae for the $D\to \bar K$ and $D_s\to \eta_s$ transition form factors are parametrized 
as~\cite{epjd4-1}
\begin{eqnarray}
     F_1^{D\to\bar K}(s) &=& \frac{0.78}{(1 - s / 2.11^2)(1 - 0.24\, s / 2.11^2)}\,, 
     \label{FF-D2K}  \\
     F_1^{D_s\to \eta_s}(s)&=& \frac{0.78}{(1 - s / 2.11^2)(1 - 0.23\, s / 2.11^2)}\,.
     \label{FF-Ds2ets}
\end{eqnarray}
Besides, we need the form factor~\cite{epjd4-1} 
\begin{eqnarray}
     A_0^{D\to \rho}(s) = \frac{0.66}{(1 - s / 1.87^2)(1 - 0.36\, s / 1.87^2)}
     \label{FF-A0-D2rho}
\end{eqnarray}
for the $D^+\to K^0_S\rho^+$ decay.  The result $F_1^{D\to\bar K}(0)=0.78$ in~\cite{epjd4-1} agree well with the lattice 
determination $F_1^{D\to\bar K}(0)= 0.765(31)$~\cite{prd96.054514}. The form factor for $D_s \to \eta$ has been measured 
by BESIII recently with the results $f^{\eta}_{+,0}(0)|V_{cs}| = 0.452\pm0.010_{\rm stat}\pm0.007_{\rm syst}$ \cite{prl132.091802}, and $f^\eta_1(0)|V_{cs}| = 0.4519\pm0.0071_{\rm stat}\pm0.0065_{\rm syst}$~\cite{prd108.092003} 
in the $D^+_s\to \eta^{(\prime)} \mu^+\nu_\mu$ and $D_s^+ \to \eta^{(\prime)} e^+ \nu_e$ decays, respectively.
In Ref.~\cite{prd110.072017}, the corresponding form factors were determined to be 
$f^{\eta}_+(0) = 0.442 \pm 0.022_{\rm stat} \pm 0.017_{\rm syst}$. Given the mixing angle 
$\phi$~\cite{prd108.092003} and $|V_{cs}|$~\cite{PDG-2024}, the result $F_1^{D_s\to\eta_s}(0)=0.78$ in 
Eq.~(\ref{FF-Ds2ets}) is in consistent with the measurements in Refs.~\cite{prd108.092003,prl132.091802,
prd110.072017} presented by BESIII Collaboration.

In the numerical calculation, we adopt the mean lives $\tau_{D^\pm}=(1033\pm5)\times 10^{-15}$ s, 
$\tau_{D^0}=(410.3\pm1.0)\times 10^{-15}$ s and $\tau_{D^\pm_s}=(501.2\pm2.2)\times 10^{-15}$ s for  $D_{(s)}$ 
mesons~\cite{PDG-2024}. The decay constants for $\rho(770)$ and its excited state $\rho(1450)$ used in this work 
are $f_{\rho(770)}=0.216\pm0.003$ GeV~\cite{jhep1608-098} and  $f_{\rho(1450)}=0.185^{+0.030}_{-0.035}$ 
GeV~\cite{plb763-29,zpc62-455}. The masses for particles in concern decays, the decay constants for kaon and pion, the 
full widths for resonances $\rho(770)$ and $\rho(1450)$ (in units of GeV), the CKM matrix elements $|V_{ud}|$ and 
$|V_{cs}|$~\cite{PDG-2024}, and the decay constants $f_{D}$ and $f_{D_s}$ for $D_{(s)}$~\cite{prd98.074512,
epjc80-113} are presented in Table~\ref{tab1}.

\begin{table*}[!]
\centering   
\caption{The branching fractions for the concerned quasi-two-body $D_{(s)}$ decays 
                with the subprocess $\rho(770)^+\to\pi^+\pi^0$, and the corresponding 
               two-body data from {\it Review of Particle Physics}~\cite{PDG-2024}.}
\label{res-R2pipi}
\setlength{\tabcolsep}{20pt}
\begin{tabular}{l c c }\hline\hline
                        \;\;\;Decay modes                                               &     $\mathcal{B}$      
                                                                                                               & Data~\cite{PDG-2024}           \\         \hline       
  $D_s^+\to\eta [\rho(770)^+\to]\pi^+\pi^0$                     &  $(7.11\pm0.20\pm0.09\pm0.47)\%$           
                                                                                                               & $(8.9\pm0.8)\%$                         \\
  $D^+\to K^0_S[\rho(770)^+\to]\pi^+\pi^0$                     & $(3.67\pm0.10\pm0.05)\%$                    
                                                                                                               & $(6.14^{+0.60}_{-0.35})\%$        \\  
  $D^0\to K^-[\rho(770)^+\to]\pi^+\pi^0$                           & $(9.12\pm0.25\pm0.11)\%$                    
                                                                                                                &  $(11.2\pm0.7)\%$                         \\  
 \hline\hline   
\end{tabular}
\end{table*}

To verify the reliability of the parameters used in this work, we calculate the branching ratios for the quasi-two-body decays, involving $D_s^+\to\eta[\rho(770)^+\to]\pi^+\pi^0$, 
$D^+\to K^0_S[\rho(770)^+\to]\pi^+\pi^0$, and $D^0\to K^-[\rho(770)^+\to]\pi^+\pi^0$, and compare them with the experimental results in Ref.~\cite{PDG-2024}.
Utilizing the differential branching fraction of Eq.~(\ref{PDW-Q2B-pi}), it is trivial to obtain the branching 
fractions for the quasi-two-body decays $D_s^+\to\eta[\rho(770)^+\to]\pi^+\pi^0$, 
$D^+\to K^0_S[\rho(770)^+\to]\pi^+\pi^0$ and $D^0\to K^-[\rho(770)^+\to]\pi^+\pi^0$ as shown in Table
\ref{res-R2pipi}, as well as the corresponding two-body data from {\it Review of Particle Physics}~\cite{PDG-2024}.
For our numerical results of the relevant quasi-two-body in Table~\ref{res-R2pipi}, the first source of the error
corresponds to the uncertainties of the decay constant $f_{\rho(770)}=0.216\pm0.003$ GeV~\cite{jhep1608-098}, 
the uncertainties of CKM matrix elements $|V_{ud}|$ and $|V_{cs}|$ in Table~\ref{tab1} contribute to the second 
source of error. One can find that these errors are quite small when comparing with their corresponding central values. 
For the $D_s$ decay, the uncertainty of mixing angle $\phi$ contributes the third error. 
We neglect the errors arising from the uncertainties of other parameters due to the tiny contributions. In view of 
$\mathcal{B}(\rho^+\to \pi^+\pi^0)\approx100\%$~\cite{PDG-2024}, our results in Table~\ref{res-R2pipi} for the decays 
$D_s^+\to\eta [\rho(770)^+\to]\pi^+\pi^0$ and $D^0\to K^-[\rho(770)^+\to]\pi^+\pi^0$ are in consistent with the data. 
However, for the decay $D^+\to K^0_S[\rho(770)^+\to]\pi^+\pi^0$, our result deviates from the experimental measurement by more than $4\sigma$. One should note that our results for the $D^+$ decay agrees with that in Ref.~\cite{prd89.054006}. This indicates that further investigations should be required to understand the branching ratio for $D^+\to K^0_S \pi^+\pi^0$.

\begin{table*}[!] 
\centering    
\caption{The predicted branching fractions for the relevant quasi-two-body $D_{(s)}$ decays with the subprocesses 
                $\rho(770,1450)^+\to \omega\pi^+$\! along with the experimental measurements in~\cite{PDG-2024} for the related three-body
                decay processes.}
\label{res-R2Opi}
\setlength{\tabcolsep}{20pt}
\begin{tabular}{l c }\hline\hline
                        \;\;\;Decay modes                                                    &  $\mathcal{B}$                                           \\         \hline       
  $D_s^+\to\eta [\rho(770)^+\to]\omega\pi^+$                      &  $(1.90\pm0.35\pm0.05\pm0.14)\times10^{-3}$      \\
  $D_s^+\to\eta [\rho(1450)^+\to]\omega\pi^+$                     &  $(0.39\pm0.16\pm0.14\pm0.03)\times10^{-3}$      \\
  $D_s^+\to\eta [\rho(770\&1450)^+\to]\omega\pi^+$          &  $(2.89\pm0.63\pm0.29\pm0.20)\times10^{-3}$      \\
  \hline  
  $D^+\to K^0_S[\rho(770)^+\to]\omega\pi^+$                     & $(1.03\pm0.20\pm0.03)\times10^{-3}$                       \\  
  $D^+\to K^0_S[\rho(1450)^+\to]\omega\pi^+$                    & $(0.16\pm0.07\pm0.06)\times10^{-3}$                       \\    
  $D^+\to K^0_S[\rho(770\&1450)^+\to]\omega\pi^+$       & $(1.53\pm0.34\pm0.15)\times10^{-3}$                       \\     
  \hline
  $D^0\to K^-[\rho(770)^+\to]\omega\pi^+$                           & $(1.96\pm0.38\pm0.05)\times10^{-3}$                       \\  
  $D^0\to K^-[\rho(1450)^+\to]\omega\pi^+$                          & $(0.28\pm0.12\pm0.10)\times10^{-3}$                       \\   
  $D^0\to K^-[\rho(770\&1450)^+\to]\omega\pi^+$              & $(2.86\pm0.63\pm0.31)\times10^{-3}$                       \\       
 \hline
                \;\;Three-body modes                                                &  Data~\cite{PDG-2024}                \\         \hline   
   $D_s^+\to\eta \omega\pi^+$                                                  & $(5.4\pm1.3)\times10^{-3}$                       \\  
  $D^+\to K^0_S \omega\pi^+$                                                 & $(7.1\pm0.5)\times10^{-3}$                       \\   
  $D^0\to K^- \omega\pi^+$                                                       & $(3.39\pm0.10)\times10^{-2}$                       \\       
 \hline\hline   
\end{tabular}
\end{table*}

Now we investigate the contributions of $\rho(770,1450)\to \omega\pi$ to relevant quasi-two-body $D_{(s)}$ 
decays. With the help of the Eq.~(\ref{PDW-Q2B-piO}), we obtain the branching fractions for the 
decays $D_s^+\to\eta[\rho^+\to]\omega\pi^+$,  $D^+\to K^0_S[\rho^+\to]\omega\pi^+$  
and $D^0\to K^-[\rho^+\to]\omega\pi^+$ in Table~\ref{res-R2Opi}, with the intermediate 
$\rho^+\in \{\rho(770)^+, \rho(1450)^+, \rho(770)^+\&\rho(1450)^+\}$. For these predicted branching fractions 
in Table~\ref{res-R2Opi}, the first source of the error corresponds to the uncertainties of 
$g_{\rho\omega\pi}=16.0\pm2.0$~GeV$^{-1}$ and $A_1=0.171\pm0.036$, the second error comes from the uncertainties 
of the decay constants for $\rho(770)$ and $\rho(1450)$, the mass $m_{\rho(1450)}=1.465\pm0.025$~GeV and full 
width $\Gamma_{\rho(1450)}=0.400\pm0.060$ GeV for $\rho(1450)$. For those three decay channels with the 
meson $\eta$ in their final states in Table~\ref{res-R2Opi},  the uncertainty of the mixing angle $\phi$ will contribute 
the third error. There are other errors for the predictions, which come from the uncertainties of masses and decay constants 
of the initial and final states, from the uncertainties of the CKM matrix elements, etc., are very small and have been neglected.

The parameters $\chi^{E,A}_q$, $\phi^{E,A}_q$, $\chi^C_V$ and $\phi^C_V$ in Eqs.~(\ref{eq-M2B})-(\ref{eq-A2B})
were fitted for the two-body $D$ decays with a light pseudoscalar and a ground vector meson as their final states.
In principle these parameters are not really appropriate to the decays with $\rho(1450)$ as the intermediate state.
But one should note that, these parameters do not exist in the dominated decay amplitudes for relevant decay processes 
in this work. In order to check the impact of these parameters on the branching fractions with 
the intermediate state $\rho(1450)$, we take $D_s^+\to\eta [\rho(1450)^+\to]\omega\pi^+$ as an example. When we vary 
the related parameters with the $50\%$ uncertainties of their values for this decay channel, we obtain its branching 
fraction as $\mathcal{B}=(0.39\pm0.03)\times10^{-3}$. Which means the effects of these parameters would be tiny by 
comparing it with the corresponding errors in Table~\ref{res-R2Opi} for $D_s^+\to\eta [\rho(1450)^+\to]\omega\pi^+$ decay.
The form factor $A_0^{D\to \rho}$ in Eq.~(\ref{FF-A0-D2rho}) is for the $D\to \rho(770)$ transition. While for the 
decay $D^+\to K^0_S[\rho(1450)^+\to]\omega\pi^+$, the form factor $A_0$ for $D\to \rho(1450)$ transition is needed.
According to the discussions in Ref.~\cite{plb763-29} with the same intermediate state $\rho(1450)$ in quasi-two-body 
$B$ meson decays, we adopt $A_0^{D\to \rho(1450)}\approx f_{\rho(1450)}/f_{\rho(770)}A_0^{D\to \rho(770)}$,  in 
view of the lack of the form factor $A_0^{D\to \rho(1450)}$ in literature. When we invest this form factor a $20\%$ 
uncertainty of its central value, we will have the branching fraction $(0.16\pm0.03)\times10^{-3}$ for the 
$D^+\to K^0_S[\rho(1450)^+\to]\omega\pi^+$ decay. Apparently, the error is not large.

In Ref.~\cite{prd107.052010}, the branching fraction for the tree-body decay $D_s^+\to\eta\omega\pi^+$ was measured 
to be $(0.54\pm0.12_{\rm stat}\pm0.04_{\rm syst})\%$; by comparing this data with the corresponding prediction 
$(2.89\pm0.72)\times10^{-3}$ in Table~\ref{res-R2Opi} for $D_s^+\to\eta [\rho(770\&1450)^+\to]\omega\pi^+$, we 
conclude that the tree-body decay $D_s^+\to\eta\omega\pi^+$ is dominated by the contribution from the subprocess 
$\rho(770\&1450)^+\to\omega\pi^+$, roughly contributing to half of the total branching fraction when employing 
the phase difference $\phi_1=\pi$ between the intermediate states $\rho(770)$ and $\rho(1450)$~\cite{plb486-29,plb562-173,
JETPL94-734,prd61.072003,prd88.054013,prd94.112001}. But in Ref.~\cite{JHEP2401-047}, with the choice of $\phi_1=\pi$, 
the shapes of predicted differential branching fractions for $B^0 \to {D}^{*-}\rho^+\to {D}^{*-}\omega\pi^+$ were found  
do not agree very well with the distribution of $\omega\pi$ measured by Belle Collaboration in~\cite{prd92.012013} for 
$B^0 \to  {D}^{*-}\omega\pi^+$ decay. The interference between the subprocesses $\rho(770)^+\to\omega\pi^+$ and 
$\rho(1450)^+\to\omega\pi^+$ is seriously affected by the phase difference $\phi_1=\pi$ for the $D_s^+\to\eta\omega\pi^+$ 
decay. We switch the phase difference $\phi_1$ from zero to $2\pi$ and find that we will obtain the maximum branching 
fraction $\mathcal{B}=3.67\times10^{-3}$ as the central value for  $D_s^+\to\eta [\rho(770\&1450)^+\to]\omega\pi^+$
when we choose $\phi_1=1.35\pi$, then the prediction is roughly $68\%$ of the measured total branching fraction for 
$D_s^+\to\eta\omega\pi^+$ presented by BESIII in~\cite{prd107.052010}.

\begin{figure}[tbp]  
\centerline{\epsfxsize=10.0cm \epsffile{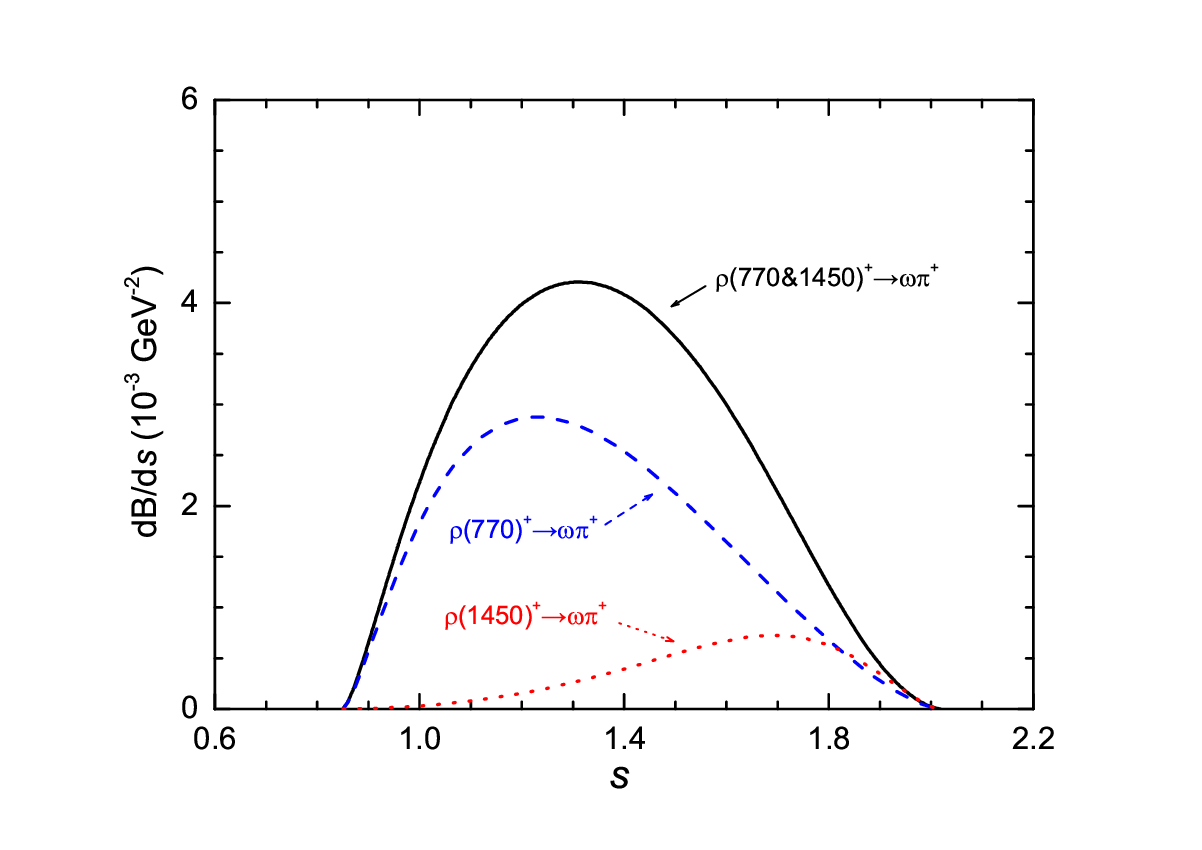}}
\vspace{-0.5cm}
\caption{The differential branching fractions for the quasi-two-body decays 
                $D_s^+\to\eta [\rho^+\to]\omega\pi^+$, with the intermediate
                $\rho^+= \{\rho(770)^+, \rho(1450)^+, \rho(770)^+\&\rho(1450)^+\}$.}
\label{fig-diff-Br}
\vspace{-0.2cm}
\end{figure}

One could argue that as a virtual bound state~\cite{Dalitz62,plb25-294} the intermediate state $\rho(770)^+$ could not 
completely exhibit its properties in the quasi-two-body decay $D_s^+\to\eta [\rho(770)^+\to]\omega\pi^+$. It is true that 
the invariant masses of the $\omega\pi$ pair exclude the region around the $\rho(770)$ pole mass. However, as we show 
in the Fig.~\ref{fig-diff-Br} the width of $\rho(770)$ renders its contribution quite sizable in the energy region of interest
of three-body $D_{(s)}$ decays. It is important to consider explicitly the subthreshold resonances, even if they contribute 
via the tail of their energy distribution, in the amplitude analysis of experimental studies.
The differential branching fractions shown in Fig.~\ref{fig-diff-Br} for the quasi-two-body decay
$D_s^+\to\eta [\rho^+\to]\omega\pi^+$ with $\rho^+= \{\rho(770)^+, \rho(1450)^+, \rho(770)^++\rho(1450)^+\}$ 
are not typical curves like $\rho(770,1450)\to \pi\pi$ in three-body $B$ meson decays~\cite{plb763-29}. The peaks of 
the dash line for $\rho(770)^+\to\omega\pi^+$ and dot line for  $\rho(1450)^+\to\omega\pi^+$ in Fig.~\ref{fig-diff-Br} 
arise from BW expressions for the involved resonances and mainly from the phase space factor in Eq.~(\ref{PDW-Q2B-piO}).

The absolute branching fraction for the three-body decay $D^+\to K^0_S\pi^+\omega$ was determined to be 
$(0.707 \pm 0.041_{\rm stat} \pm 0.029_{\rm syst})\%$ by BESIII in Ref.~\cite{prd105.032009}, which is $4.6$ times 
larger than the prediction $(1.53\pm0.37)\times10^{-3}$ in Table~\ref{res-R2Opi} for the quasi-two-body decay 
$D^+\to K^0_S[\rho(770\&1450)^+\to]\omega\pi^+$. We need to stress that the contribution weight of 
$D^+\to K^0_S[\rho(770\&1450)^+\to]\omega\pi^+$ in the corresponding three-body decay could be enhanced 
by improving the calculation method employed in this work, since our prediction for 
$D^+\to K^0_S[\rho(770)^+\to]\pi^+\pi^0$ in Table~\ref{res-R2pipi} is just around half of the data in 
\cite{PDG-2024}. Which also means the branching fraction for $D^+\to K^0_S[\rho(770\&1450)^+\to]\omega\pi^+$ 
could be twice of the corresponding prediction in Table~\ref{res-R2Opi}. 
In addition, the $D^+\to b_1(1235)$ transition with a emitted $K^0_S$, could generate the 
$\omega\pi^+$ pair from $b_1(1235)$ for $D^+\to K^0_S\pi^+\omega$ decay. The $\omega\pi^+$ pair generated 
from the $D^+\to b_1(1235)$ transition is not like it from the matrix element 
$\langle\omega\pi|V^\mu\!-\!A^\mu|0\rangle$ in $D_s^+\to\eta \omega\pi^+$ decay, whose 
amplitude is just proportional to the mass difference between $u$ and $d$ quarks. The measurement of 
${\mathcal B}(D^+\to b_{1}(1235)^0e^{+}\nu_{e})\times {\mathcal B} (b_1(1235)^0\to \omega \pi^0) = 
(1.16\pm0.44_{\rm stat}\pm0.16_{\rm syst})\times10^{-4}$ by BESIII~\cite{ar2407-20551} hinting possible contribution 
of $b_1(1235)^+\to \omega \pi^+$ for $D^+\to K^0_S\pi^+\omega$ decay.
But we know the contribution shouldn't be large since we have the theoretical estimation $1.7\times 10^{-3}$ as 
the branching fraction for $D^+\to \bar{K}^0 b_1(1235)^+$~\cite{prd67.094007}.

The input parameters for the quasi-two-body 
$D$ meson decays were updated in Ref.~\cite{ar2503.18593} very recently. With those updated inputs 
in~\cite{ar2503.18593},  the quasi-two-body decay branching fractions for the $D_s^+\to\eta \rho(770)^+$ 
and $D^0\to K^-\rho(770)^+$ agree well with these corresponding results in Tables~\ref{res-R2pipi}-\ref{res-R2Opi}
with the subprocesses $\rho(770)^+\to\pi^+\pi^0$ and $\rho(770)^+\to\omega\pi^+$, respectively.  
While the branching fractions for the quasi-two-body decays $D^+\to K^0_S[\rho(770)^+\to]\pi^+\pi^0$ and 
$D^+\to K^0_S[\rho(770\&1450)^+\to]\omega\pi^+$ will be increased to $(4.45\pm0.14)\%$ and 
$(2.63\pm0.64)\times 10^{-3}$, respectively. And the latter numerical value is about $37\%$ of the total branching 
fraction of $D^+\to K^0_S\pi^+\omega$ in~\cite{prd105.032009}. For this three-body decay 
$D^+ \to K^0_S\pi^+\omega$, one also has the contribution from $D^+ \to \bar{K}^{*0}\pi^+$ with the subprocess 
$\bar{K}^{*0}\to K^0_S\omega$. When we take into account the branching fraction $(1.04\pm0.12)\%$ for 
$D^+ \to \bar{K}^{*0}\pi^+$ along with $\bar{K}^{*0}\to K^-\pi^+$~\cite{PDG-2024} and the discussions 
for the virtual contributions of $K^*$ in~\cite{2508.09578}, we can't expect a large contribution from 
$D^+ \to \bar{K}^{*0}\pi^+\to  K^0_S\omega \pi^+$ for this three-body decay process. In addition, the cascade 
decay $D^+\to K^*(892)^+\omega\to K^0_S\pi^+\omega$ is a doubly Cabibbo suppressed model. Then we are pretty 
sure that the quasi-two-body process $D^+\to K^0_S[\rho(770\&1450)^+\to]\omega\pi^+$ should be the 
most important one for the three-body decay $D^+ \to K^0_S\pi^+\omega$.

The branching fraction for the decay $D^0\to K^-[\rho(770\&1450)^+\to]\omega\pi^+$ is predicted to be 
$(2.86\pm0.70)\times10^{-3}$ in this work, which is less than a tenth of the measurement 
$(3.392 \pm 0.044_{\rm stat} \pm 0.085_{\rm syst})\%$ for three-body decay $D^0\to K^-\pi^+\omega$ in 
Ref.~\cite{prd105.032009}. The data $(6.5\pm3.0)\times10^{-3}$ for $D^0\to \bar{K}^*(892)^0\omega$ with 
the subprocesses  $\bar{K}^*(892)^0 \to K^-\pi^+$ and $\omega \to \pi^+\pi^-\pi^0$ is about twice of our result 
for $D^0\to K^-[\rho(770\&1450)^+\to]\omega\pi^+$ in this work. 
Actually, the three-body decay $D^0\to K^-\pi^+\omega$ is very different from the process 
 $D_s^+\to\eta \omega\pi^+$, the former has very rich intermediate states. One has the resonances 
$\bar{K}^*(892,1410,1680)^0, \bar{K}_{0,2}^*(1430)^0$, ${K}^*(892,1410,1680)^-, {K}_1(1270)^-$ 
and $\rho(770,1450)^+$ decaying into $K^-\pi^+$, $K^- \omega$ and $\pi^+\omega$, respectively, in this
three-body decay $D^0\to K^-\pi^+\omega$. The analyses of the complete resonance contributions for the decay 
process $D^0\to K^-\pi^+\omega$ is beyond the scope of this work, we shall leave them to the future studies.

\section{Summary}
\label{sec-sum}

The Cabibbo-favored three-body decays $D_s^+\to \eta \omega\pi^+$, $D^+\to K^0_S\pi^+\omega$ and 
$D^0\to K^-\pi^+\omega$ have been observed by BESIII Collaboration recently, but without any amplitude analysis. 
In order to understand better the relevant experimental measurements, we studied the contributions of the subprocesses 
$\rho(770,1450)^+\to \omega\pi^+$ in these three-body $D$ meson decays by introducing them into the decay amplitudes 
of the relevant decay channels via the vector form factor $F_{\omega\pi}(s)$, which has been measured in the related 
processes of $\tau$ decay and $e^+e^-$ annihilation. 

With the parameters $g_{\rho\omega\pi}=16.0\pm2.0$ GeV$^{-1}$ and $A_1=0.171\pm0.036$ for the 
vector form factor $F_{\omega\pi}$, we predicted the branching fractions for the first time for the
quasi-two-body decays $D_s^+\to\eta[\rho^+\to]\omega\pi^+$,  $D^+\to K^0_S[\rho^+\to]\omega\pi^+$ and 
$D^0\to K^-[\rho^+\to]\omega\pi^+$ with the intermediate $\rho^+ = \{\rho(770)^+, \rho(1450)^+, \rho(770)^++\rho(1450)^+\}$. By comparing our predictions with the experimental data, we found that the contributions from the 
subprocess $\rho(770)^+\to\omega\pi^+$ are  significant in these three-body decays $D_s^+\to\eta \omega\pi^+$, 
$D^+\to K^0_S \omega\pi^+$ and $D^0\to K^- \omega\pi^+$, notwithstanding the contributions originating from 
the Breit-Wigner tail effect of $\rho(770)^+$. The interference effects between the resonances $\rho(770)^+$ and 
$\rho(1450)^+$ make the $P$-wave resonance contributions for $\omega\pi^+$ pair enhanced in these three-body decays. 
The numerical results of this study suggest that the dominant resonance contributions for the three-body decays 
$D_s^+\to\eta \omega\pi^+$ and $D^+\to K^0_S \omega\pi^+$ are originated from the $P$-wave intermediate states 
$\rho(770)^+$, $\rho(1450)^+$ decaying into $\omega\pi^+$ pair and their interference effects. 

\begin{acknowledgments}
This work was supported in part by the National Natural Science Foundation of China under Grant 
No. 12465017,  the Foundation of Shanxi ``1331 Project" Key Subjects Construction.
\end{acknowledgments}





\begin{thebibliography}{99}
\addtolength{\itemsep}{0.2ex}

\bibitem{prd107.052010}
M.~Ablikim \textit{et al.} (BESIII Collaboration), 
Phys. Rev. D \textbf{107},  052010 (2023).

\bibitem{prd80.051102}
J.~Y.~Ge \textit{et al.} (CLEO Collaboration), 
Phys. Rev. D \textbf{80}, 051102 (2009).

\bibitem{PDG-2024}
S.~Navas \textit{et al.}  (Particle Data Group), 
Phys. Rev. D \textbf{110}, 030001 (2024).
  
\bibitem{prl123.112001} 
M.~Ablikim \textit{et al.} (BESIII Collaboration),
Phys. Rev. Lett. \textbf{123},  112001 (2019).

\bibitem{epjc80-895}
Y.~K.~Hsiao, Y.~Yu, and B.~C.~Ke,
Eur. Phys. J. C \textbf{80}, 895 (2020).

\bibitem{plb803-135279}
R.~Molina, J.~J.~Xie, W.~H.~Liang, L.~S.~Geng, and E.~Oset,
Phys. Lett. B \textbf{803}, 135279 (2020).

\bibitem{epjc80-1041}
M.~Y.~Duan, J.~Y.~Wang, G.~Y.~Wang, E.~Wang, and D.~M.~Li,
Eur. Phys. J. C \textbf{80}, 1041 (2020).

\bibitem{prd103.116016}
X.~Z.~Ling, M.~Z.~Liu, J.~X.~Lu, L.~S.~Geng, and J.~J.~Xie,
Phys. Rev. D \textbf{103}, 116016 (2021).

\bibitem{prd109.076027}
M.~Bayar, R.~Molina, E.~Oset, M.~Z.~Liu, and L.~S.~Geng,
Phys. Rev. D \textbf{109}, 076027 (2024).

\bibitem{epjc81-1093}
Y.~Yu, Y.~K.~Hsiao, and B.~C.~Ke,
Eur. Phys. J. C \textbf{81}, 1093 (2021), 
arXiv:2108.02936 [hep-ph].

\bibitem{zpc57-495}
R.~Decker and E.~Mirkes,
Z. Phys. C \textbf{57}, 495 (1993).

\bibitem{prd86.037302}
N.~Paver and Riazuddin,
Phys. Rev. D \textbf{86}, 037302 (2012).

\bibitem{pr112.1375}
S.~Weinberg,
Phys. Rev. \textbf{112}, 1375-1379 (1958)

\bibitem{ar0807.4900}
B.~Aubert \textit{et al.} (BaBar Collaboration), 
arXiv:0807.4900 [hep-ex].

\bibitem{npb218-110}
K.~E.~Alwyn (BaBar Collaboration),
Nucl. Phys. B (Proc. Suppl.) \textbf{218}, 110  (2011).

\bibitem{BW-model}  
  G.~Breit and E.~Wigner,
  Phys.\ Rev.\  {\bf 49}, 519 (1936).
 
\bibitem{Dalitz62} 
  R.~H.~Dalitz,
  {\it Strange particles and strong interactions,}
  (Oxford Press, London, 1962).  
  
\bibitem{plb25-294}  
  A.~Astier, J.~Cohen-Ganouna, M.~D.~Negra, B.~Mar\'{e}chal, L.~Montanet, M.~Tomas, M.~Baubillier, and J.~Duboc,
  Phys. Lett. B \textbf{25}, 294 (1967).
  
\bibitem{prd94.072001}  
  R.~Aaij {\it et al.} (LHCb Collaboration),
  Phys.\ Rev.\ D {\bf 94}, 072001 (2016).
  
\bibitem{plb791-342}  
  W.~F.~Wang and J.~Chai,
  Phys.\ Lett.\ B {\bf 791}, 342 (2019).
  
\bibitem{plb185-223} 
H.~Albrecht \textit{et al.} (ARGUS Collaboration), 
Phys. Lett. B \textbf{185}, 223 (1987).

\bibitem{prd61.072003} 
K.~W.~Edwards \textit{et al.} (CLEO Collaboration), 
Phys. Rev. D \textbf{61}, 072003 (2000).

\bibitem{rmp78.1043} 
M.~Davier, A.~H\"ocker, and Z.~Zhang,
Rev. Mod. Phys. \textbf{78}, 1043 (2006).

\bibitem{prl103.041802} 
B.~Aubert \textit{et al.} (BaBar Collaboration), 
Phys. Rev. Lett. \textbf{103}, 041802 (2009).

\bibitem{2309.00280} 
M.~N.~Achasov \textit{et al.} (SND Collaboration), 
Phys. Rev. D \textbf{108}, 092012 (2023).

\bibitem{prd88.054013} 
M.~N.~Achasov \textit{et al.} (SND Collaboration), 
Phys. Rev. D \textbf{88}, 054013 (2013).

\bibitem{plb562-173} 
R.~R.~Akhmetshin \textit{et al.} (CMD-2 Collaboration), 
Phys. Lett. B \textbf{562}, 173 (2003),
arXiv:hep-ex/0304009 [hep-ex].

\bibitem{plb174-453} 
S.~I.~Dolinsky \textit{et al.}, 
Phys. Lett. B \textbf{174}, 453 (1986).

\bibitem{plb486-29} 
M.~N.~Achasov \textit{et al.} (SND Collaboration), 
Phys. Lett. B \textbf{486}, 29 (2000).

\bibitem{plb669-223} 
F.~Ambrosino \textit{et al.} (KLOE Collaboration), 
Phys. Lett. B \textbf{669}, 223 (2008).

\bibitem{JETPL94-734} 
M.~N.~Achasov \textit{et al.} (SND Collaboration), 
JETP Lett. \textbf{94}, 734  (2012).

\bibitem{prd94.112001} 
M.~N.~Achasov \textit{et al.} (SND Collaboration), 
Phys. Rev. D \textbf{94}, 112001 (2016).

\bibitem{prd96.092009} 
J.~P.~Lees \textit{et al.} (BaBar Collaboration), 
Phys. Rev. D \textbf{96}, 092009 (2017).

\bibitem{plb813-136059} 
M.~Ablikim \textit{et al.} (BESIII Collaboration),  
Phys. Lett. B \textbf{813}, 136059 (2021).

\bibitem{prd64.092001} 
J.~P.~Alexander \textit{et al.} (CLEO Collaboration),
Phys. Rev. D \textbf{64}, 092001 (2001).

\bibitem{prd74.012001} 
B.~Aubert \textit{et al.} (BaBar Collaboration),
Phys. Rev. D \textbf{74}, 012001 (2006).

\bibitem{prd92.012013}
D.~Matvienko \textit{et al.} (Belle Collaboration), 
Phys. Rev. D \textbf{92},  012013 (2015).

\bibitem{prd60.114011}  
A.~Donnachie and Y.~S.~Kalashnikova,
Phys. Rev. D \textbf{60}, 114011 (1999).

\bibitem{npb443-233} 
F.~E.~Close and P.~R.~Page,
Nucl. Phys. B \textbf{443}, 233 (1995);

\bibitem{prd52.1706} 
F.~E.~Close and P.~R.~Page,
Phys. Rev. D \textbf{52}, 1706 (1995);

\bibitem{prd56.1584} 
F.~E.~Close and P.~R.~Page,
Phys. Rev. D \textbf{56}, 1584 (1997).

\bibitem{prd55.4157} 
T.~Barnes, F.~E.~Close, P.~R.~Page, and E.~S.~Swanson,
Phys. Rev. D \textbf{55}, 4157 (1997).

\bibitem{prd105.032009}
M.~Ablikim \textit{et al.} (BESIII Collaboration), 
Phys. Rev. D \textbf{105}, 032009 (2022).

\bibitem{pr135.B551} 
 G.~N.~Fleming,
 Phys.\ Rev.  {\bf 135}, B551 (1964).
  
\bibitem{pr166.1731} 
 D.~Morgan,
 Phys.\ Rev.  {\bf 166}, 1731 (1968).
  
\bibitem{prd11.3165} 
 D.~Herndon, P.~Soding, and R.~J.~Cashmore, 
 Phys.\ Rev.\ D {\bf 11}, 3165 (1975).

\bibitem{JHEP2401-047}
Y.~S.~Ren, A.~J.~Ma, and W.~F.~Wang,
JHEP \textbf{01}, 047 (2024).

\bibitem{prd101.111901}  
  W.~F.~Wang,
  Phys.\ Rev.\ D {\bf 101,} 111901(R) (2020); arXiv:2004.09027[hep-ph]. 

\bibitem{prd103.056021} 
  W.~F.~Wang,
  Phys. Rev. D \textbf{103}, 056021 (2021).

\bibitem{prd103.016002}  
  A.~J.~Ma and W.~F.~Wang,
  Phys. Rev. D \textbf{103}, 016002 (2021).
    
\bibitem{cpc46-053104}  
 A.~J.~Ma and W.~F.~Wang,
 Chin. Phys. C \textbf{46}, 053104 (2022).
 
\bibitem{prd107.116023}
S.~H.~Zhou, X.~X.~Hai, R.~H.~Li, and C.~D.~L\"u,
Phys. Rev. D \textbf{107}, 116023 (2023).

\bibitem{prd110.056001}
S.~H.~Zhou, R.~H.~Li, and X.~Y.~L\"u,
Phys. Rev. D \textbf{110}, 056001 (2024).
 
\bibitem{prd109.116009}
W.~F.~Wang, L.~F.~Yang, A.~J.~Ma, and \`A.~Ramos,
Phys. Rev. D \textbf{109}, 116009 (2024).

\bibitem{prd104.116019}  
 W.~F.~Wang,
 Phys. Rev. D \textbf{104}, 116019 (2021).

\bibitem{prd85.092016} 
  J.~Insler {\it et al.} (CLEO Collaboration),
  Phys.\ Rev.\ D {\bf 85}, 092016 (2012); {\bf 94}, 099905(E) (2016).
 
\bibitem{prd93.052018}  
  R.~Aaij {\it et al.} (LHCb Collaboration),
  Phys.\ Rev.\ D {\bf 93}, 052018 (2016).
 
\bibitem{prd103.114028}  
  J.-P.~Dedonder, R.~Kami\'nski, L.~Le\'sniak, and B.~Loiseau,
  Phys. Rev. D \textbf{103}, 114028 (2021).
  
\bibitem{prd105.074035} 
Q.~S.~Zhou, J.~Z.~Wang, X.~Liu, and T.~Matsuki,
Phys. Rev. D \textbf{105}, 074035 (2022).

\bibitem{prd85.054014}
B.~Bhattacharya, M.~Gronau, and J.~L.~Rosner,
Phys. Rev. D \textbf{85}, 054014 (2012).
 
\bibitem{prd86.014014}
H.~Y.~Cheng and C.~W.~Chiang,
Phys. Rev. D \textbf{86}, 014014 (2012).

\bibitem{prd85.034036}
H.~Y.~Cheng and C.~W.~Chiang,
Phys. Rev. D \textbf{85}, 034036 (2012); \textbf{85}, 079903(E) (2012).

\bibitem{prd93.114010} 
H.~Y.~Cheng, C.~W.~Chiang, and A.~L.~Kuo,
Phys. Rev. D \textbf{93}, 114010 (2016).

\bibitem{prd100.093002}  
H.~Y.~Cheng and C.~W.~Chiang,
Phys. Rev. D \textbf{100}, 093002 (2019)

\bibitem{prd104.073003}
H.~Y.~Cheng and C.~W.~Chiang,
Phys. Rev. D \textbf{104}, 073003 (2021).

\bibitem{prd110.094052}
H.~Y.~Cheng, C.~W.~Chiang, and F.~R.~Xu,
Phys. Rev. D \textbf{110}, 094052 (2024).

\bibitem{prd86.036012}
H.~n.~Li, C.~D.~L\"u, and F.~S.~Yu,
Phys. Rev. D \textbf{86}, 036012 (2012).
 
\bibitem{prd89.054006} 
Q.~Qin, H.~n.~Li, C.~D.~L\"u, and F.~S.~Yu,
Phys. Rev. D \textbf{89}, 054006 (2014).

\bibitem{adv-7627308} 
  H.~Zhou, B.~Zheng, and Z.~H.~Zhang,
  Adv.\ High Energy Phys.\  {\bf 2018}, 7627308 (2018).
 
\bibitem{ar2503.18593}
X.~D.~Zhou and S.~H.~Zhou,
Phys. Rev. D \textbf{111}, 116008 (2025).
  
\bibitem{1605.03889}      
  J.~H.~Alvarenga Nogueira  \textit{et al.}, 
  arXiv:1605.03889 [hep-ex].

\bibitem{prd96.113003} 
  D.~Boito, J.~P.~Dedonder, B.~El-Bennich, R.~Escribano, R.~Kami\'nski, L.~Le\'sniak, and B.~Loiseau,
  Phys. Rev. D \textbf{96}, 113003 (2017).

\bibitem{rmp68.1125} 
 G.~Buchalla, A.~J.~Buras, and M.~E.~Lautenbacher,
 Rev. Mod. Phys. \textbf{68}, 1125 (1996).

\bibitem{Cabibbo}
N.~Cabibbo,
Phys. Rev. Lett. \textbf{10}, 531 (1963). 

\bibitem{Kobayashi}
M.~Kobayashi and T.~Maskawa,
Prog. Theor. Phys. \textbf{49}, 652 (1973). 

\bibitem{BW-X}  
  J.~M. Blatt and V.~F. Weisskopf,  
  {\it Theoretical nuclear physics} 
  (Springer, New York, 1952).

\bibitem{prd63.092001}  
  S.~Kopp \textit{et al.} (CLEO Collaboration),
  Phys. Rev. D \textbf{63}, 092001 (2001).

\bibitem{epjc39-41}   
  C.~Bruch, A.~Khodjamirian, and J.~H.~K\"uhn,
  Eur.\ Phys.\ J. C {\bf 39,} 41 (2005).
   
\bibitem{prd81-094014}  
  H.~Czy\.z, A.~Grzeli\'nska, and J.~H.~K\"uhn,
  Phys.\ Rev.\ D {\bf 81,} 094014 (2010).

\bibitem{plb66-165}  
D.~H.~Boal, B.~J.~Edwards, A.~N.~Kamal, R.~Rockmore, and R.~Torgerson,
Phys. Lett. B \textbf{66}, 165 (1977),
 
\bibitem{prd30.594}  
\"O.~Kaymakcalan, S.~Rajeev, and J.~Schechter,
Phys. Rev. D \textbf{30}, 594 (1984).

\bibitem{prd46.1195}  
S.~Fajfer, K.~Suruliz, and R.~J.~Oakes,
Phys. Rev. D \textbf{46}, 1195 (1992).

\bibitem{epja38-331}  
S.~Pacetti,
Eur. Phys. J. A \textbf{38}, 331 (2008).

\bibitem{prd92.014014}    
I.~Caprini,
Phys. Rev. D \textbf{92}, 014014 (2015).

\bibitem{2307.10357}  
H.~Sch\"afer, M.~Zanke, Y.~Korte, and B.~Kubis,
Phys. Rev. D \textbf{108}, 074025 (2023).

\bibitem{ppnp120-103884} 
S.~S.~Fang, B.~Kubis, and A.~Kup\'s\'c,
Prog. Part. Nucl. Phys. \textbf{120}, 103884 (2021).

\bibitem{jpg36-085008} 
G.~Li, Y.~J.~Zhang, and Q.~Zhao,
J. Phys. G \textbf{36}, 085008 (2009).

\bibitem{prc83.048201}  
A.~B.~Arbuzov, E.~A.~Kuraev and M.~K.~Volkov,
Phys. Rev. C \textbf{83}, 048201 (2011).
  
\bibitem{prd86.057301}
M.~K.~Volkov, A.~B.~Arbuzov and D.~G.~Kostunin,
Phys. Rev. D \textbf{86}, 057301 (2012). 
  
\bibitem{prd55.249}
M.~Lublinsky,
Phys. Rev. D \textbf{55}, 249 (1997).

\bibitem{prd77.113011}
A.~Flores-Tlalpa and G.~L\'opez-Castro,
Phys. Rev. D \textbf{77}, 113011 (2008).

\bibitem{prd58.114006}
T.~Feldmann, P.~Kroll, and B.~Stech,
Phys. Rev. D \textbf{58}, 114006 (1998).

\bibitem{plb449-339}
T.~Feldmann, P.~Kroll, and B.~Stech,
Phys. Lett. B \textbf{449}, 339 (1999).

\bibitem{JHEP0907-105}
F.~Ambrosino \textit{et al.} (KLOE Collaboration),
JHEP \textbf{07}, 105 (2009).
 
\bibitem{plb648-267}
F.~Ambrosino \textit{et al.} (KLOE Collaboration),
Phys. Lett. B \textbf{648}, 267 (2007).

\bibitem{JHEP1501-024}
R.~Aaij \textit{et al.} (LHCb Collaboration),
JHEP \textbf{01}, 024 (2015).

\bibitem{prl122.121801}
M.~Ablikim \textit{et al.} (BESIII Collaboration),
Phys. Rev. Lett. \textbf{122}, 121801 (2019).

\bibitem{prd108.092003}
M.~Ablikim \textit{et al.} (BESIII Collaboration),
Phys. Rev. D \textbf{108},  092003 (2023).

\bibitem{epjd4-1}
D.~Melikhov,
\href{https://link.springer.com/article/10.1007/s1010502c0002}
 {Eur. Phys. J. direct \textbf{4}, no.1, 2 (2002)};
arXiv:hep-ph/0110087 [hep-ph].
 
\bibitem{prd96.054514}
V.~Lubicz \textit{et al.}  (ETM Collaboration),  
Phys. Rev. D \textbf{96}, 054514 (2017); \textbf{99}, 099902(E) (2019); \textbf{100}, 079901(E) (2019).

\bibitem{prl132.091802}
M.~Ablikim \textit{et al.} (BESIII Collaboration),
Phys. Rev. Lett. \textbf{132}, 091802 (2024).

\bibitem{prd110.072017}
M.~Ablikim \textit{et al.} (BESIII Collaboration),
Phys. Rev. D \textbf{110}, 072017 (2024).

\bibitem{jhep1608-098}
  A.~Bharucha, D.~M.~Straub, and R.~Zwicky,
  J. High Energy Phys. {08} (2016) 098.
    
\bibitem{plb763-29}  
  W.~F.~Wang and H.~n.~Li,
  Phys.\ Lett.\ B {\bf 763}, 29 (2016).

\bibitem{zpc62-455}
  A.~B.~Clegg and A.~Donnachie,
  Z.\ Phys. C {\bf 62,}  455 (1994).

\bibitem{prd98.074512}
A.~Bazavov \textit{et al.} (Fermilab Lattice and MILC Collaborations), 
Phys. Rev. D \textbf{98}, 074512 (2018).
    
\bibitem{epjc80-113}
  S.~Aoki \textit{et al.} (Flavour Lattice Averaging Group),
  Eur. Phys. J. C \textbf{80}, 113 (2020).

\bibitem{ar2407-20551}
M.~Ablikim \textit{et al.} (BESIII Collaborations), 
arXiv:2407.20551 [hep-ex].

\bibitem{prd67.094007}
H.~Y.~Cheng,
Phys. Rev. D \textbf{67}, 094007 (2003).

\bibitem{2508.09578}
A.~J.~Ma and W.~F.~Wang,
arXiv:2508.09578 [hep-ph].

\end{thebibliography}
\end{document}